\documentclass[aps,prc,preprint,tightenlines,showpacs,
showkeys,superscriptaddress,nofootinbib,14pt]{revtex4}

\usepackage{graphicx}
\usepackage[figuresright]{rotating}

\begin{document}


\title{Coherent Description for Hitherto Unexplained
Radioactivities by Super- and Hyperdeformed Isomeric
States}


\author{A.~Marinov}
\email{marinov@vms.huji.ac.il}
\author{S.~Gelberg}
\affiliation{The Racah Institute of Physics, The Hebrew
University, Jerusalem 91904, Israel}
\author{D.~Kolb}
\affiliation{Department of Physics, University GH Kassel, 34109
Kassel, Germany}
\author{R.~Brandt}
\affiliation{Kernchemie, Philipps University, 35041 Marburg,
Germany}
\author{A.~Pape}
\affiliation{IReS-UMR7500, IN2P3-CNRS/ULP, BP28, F-67037
Strasbourg cedex 2, France}


\date{March 9, 2003}

\begin{abstract}
Recently long-lived high spin super- and hyperdeformed isomeric states with
unusual radioactive decay properties have been
discovered. Based on these
newly observed modes of radioactive decay, consistent interpretations are
suggested for previously unexplained phenomena seen in nature. These are the Po
halos, the low-energy enhanced 4.5 MeV $\alpha$-particle
group proposed to be due to
an isotope of a superheavy element with Z = 108,
 and the giant halos.

\end{abstract}

\maketitle

\section{Introduction}
     Despite the intensive study of nuclear physics and radioactivity for more than a
century, there are still several unexplained phenomena seen in
nature. In the present paper we consider three such phenomena. The
first one is that of the Po halos observed in mica
\cite{hen39,gen68} where the concentric halos correspond to the
$\alpha$-particle decay chains of $^{210}$Po, $^{214}$Po and
$^{218}$Po.\footnote{Colored pictures of various halos are given
in Ref. \cite{gen92}.}
Since the lifetimes of these isotopes are short
and halos belonging to their long-lived precursors from the
$^{238}$U decay chain are absent, their origin is puzzling. The
second unexplained phenomenon is the observation
\cite{cher63,chery64,cher68,mei70} in several minerals, using
solid state detectors, of a low energy 4.5-MeV $\alpha$-particle
group with an estimated \cite{cher63}
 half-life of (2.5$\pm$0.5)x10$^{8}$ y
which, based on chemical behavior, has been suggested to be due to
the decay of an isotope of Eka-Os (Z = 108; Hs). However, 4.5 MeV
is a low energy compared to the predicted 9.5 - 6.7 MeV for
$\beta$-stable isotopes of Hs \cite{mol97,kou00,lir00}, and
T$_{1/2}$ = 2.5x10$^{8}$ y is too short by a factor of 10$^{8}$,
compared to predictions \cite{vio66,roy00}
 from the lifetime versus energy
relationship for normal 4.5 MeV $\alpha$-particles from Hs. Still another unexplained
phenomenon is that of the giant halos \cite{gen70}.
Halos, with radii which may correspond to 10
and 13 MeV $\alpha$-particles, have been seen in mica \cite{gen70}.
Unlike the situation with the Po
halos, here it is not absolutely certain that their origin is from such
high energy $\alpha$-particles \cite{gen70,bra78,mol80}.
However, if they are, then their existence is unexplained. For nuclei
around the $\beta$-stability valley, 10 and 13 MeV $\alpha$-particles
are respectively predicted \cite{mol97,kou00,lir00}
for Z values around 114 and 126. The estimated \cite{roy00} half-life
for 10 MeV  $\alpha$'s in Z = 114
nuclei is about 1 s, and for 13 MeV $\alpha$'s in Z = 126 nuclei, it is about 10$^{-4}$ s.
 It is not
clear how such high-energy $\alpha$-particles with such short
predicted lifetimes can exist in nature.\footnote{On the stability
of superheavy elements as predicted in the late sixties of the
previous century and on the current situation see Refs.
\cite{mos69} and \cite{gre02}, respectively.} The purpose of this
paper is to propose consistent interpretations for the three
unexplained phenomena, based on the recently discovered
\cite{mar96a,mar96b,mar01a,mar01b}
 high spin long-lived super-
and hyperdeformed isomeric states and their unconventional decay properties.
Preliminary results have been presented before \cite{mar01c,mar02}.
\section{Super- and Hyperdeformed Isomeric States and Abnormal
Radioactive decays}
In a study of the $^{16}$O + $^{197}$Au
\cite{mar96a,mar96b}
      and  $^{28}$Si + $^{181}$Ta \cite{mar01a} reactions at and below the Coulomb
barrier, unusually low energy very life-time-enhanced $\alpha$-particle
group \cite{mar96a} on the one
hand, and high energy strongly retarded $\alpha$-particles \cite{mar01a}
 on the other hand, have been
observed in coincidence with superdeformed (SD) $\gamma$-ray
transitions. In addition, long-lived proton radioactivities
\cite{mar96b,mar01a} have been found. These unusual radioactive
processes have been explained as due to long-lived high spin
super- and hyperdeformed (HD) isomeric
state.\footnote{Experimental evidence for hyperdeformed states in
U isotopes has been seen in Ref. \cite{kra98}.} The situation is
summarized schematically in Fig. 1. A SD isomeric state can decay
by emitting very enhanced $\alpha$-particles to a similar state of
the daughter nucleus, or by strongly retarded $\alpha$-particles
to a normal deformed or the ground state (g.s.) of the daughter.
It can also decay by retarded proton radioactivity. Similarly, a
HD isomeric state can decay by retarded $\alpha$-particles to SD
states, or by enhanced $\alpha$-particles to HD states of the
daughter nucleus. As mentioned above all these unusual radioactive
decays have been seen experimentally
\cite{mar96a,mar96b,mar01a,mar01b}.

\begin{widetext}

\begin{figure}
\mbox{
\begin{minipage}[h] {0.46\linewidth}
\includegraphics[height=6.6cm]{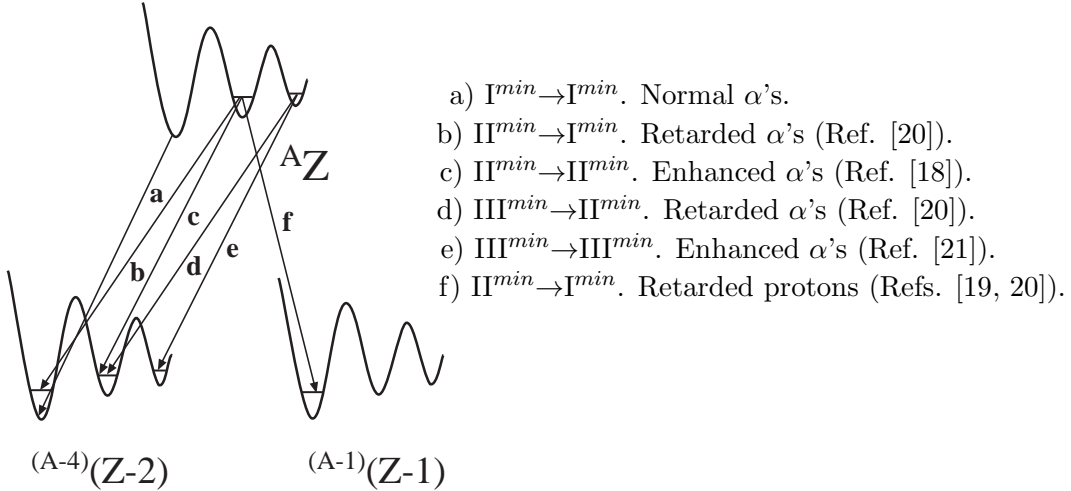}
\vspace*{0.5cm}
\end{minipage}\hfill

\begin{minipage}[h] {0.46\linewidth}
\vspace*{-2.0cm}
\hspace*{-4.8cm}a) I$^{min}$$\rightarrow$I$^{min}$. Normal $\alpha$'s.\\
\hspace*{-2.8cm}b) II$^{min}$$\rightarrow$I$^{min}$.
Retarded $\alpha$'s (Ref. \cite{mar01a}).\\
\hspace*{-2.5cm}c) II$^{min}$$\rightarrow$II$^{min}$.
Enhanced $\alpha$'s (Ref. \cite{mar96a}).\\
\hspace*{-2.5cm}d) III$^{min}$$\rightarrow$II$^{min}$.
 Retarded $\alpha$'s (Ref. \cite{mar01a}).\\
\hspace*{-2.3cm} e) III$^{min}$$\rightarrow$III$^{min}$.
 Enhanced $\alpha$'s (Ref. \cite{mar01b}).\\
\hspace*{-1.4cm} f) II$^{min}$$\rightarrow$I$^{min}$.
Retarded protons (Refs. \cite{mar96b,mar01a}).\\
\end{minipage}
}
\vspace*{-0.2cm} \caption{Summary of new types of particle decays
seen in different experiments}
\end{figure}

\end{widetext}

These isomeric states have already been used \cite{mar01b} to
interpret
     previously low energy unidentified $\alpha$-particle groups seen in actinide sources
     produced by secondary reactions in a CERN W
target \cite{mar71}, and to explain the production \cite{mar71},
in the same W targets, of a long-lived superheavy element with Z =
112, which was based on the observation of fission fragments in
separated Hg sources, and on measurements of the masses of the
fissioning nuclei \cite{mar84}.
\section{Super- and Hyperdeformed Isomeric States and the Puzzling
Phenomena Seen in Nature}
     As stated above, the discovery of these isomeric states with their unusual decay
properties enables one to also explain the previously unexplained
phenomena seen in nature.

     The Po halos \cite{hen39,gen68} may be due to similar isomeric states in
     nuclei with Z $\approx$ 84 which
eventually decayed by $\beta$- or $\gamma$-decays to the g.s. of
$^{210}$Po, $^{214}$Po and $^{218}$Po.

The observed \cite{cher63,chery64,cher68,mei70}
      4.5-MeV $\alpha$-particle group might be due to an
      enhanced III$^{min}$ $\rightarrow$ III$^{min}$ (HD $\rightarrow$ HD) transition
       in a Z = 108, A $\approx$ 270 nucleus. By using various possible
deformation parameters for the third minimum, the predicted
\cite{mar01b} half-life for such a transition, as seen in Table 1,
is around 10$^{9}$ y. This value is in reasonable agreement with
the experimental estimate \cite{cher63} of around 2.5x10$^{8}$ y
and it differs greatly from the predicted value \cite{vio66,roy00}
of about 5x10$^{16}$ y for a normal 4.5-MeV $\alpha$-transition
 from Z = 108.

\begin{table}[h]
\caption{Calculated half-lives for hyperdeformed to hyperdeformed
$\alpha$-particle transition of 4.5 MeV from $^{271}$Hs assuming
various deformation parameters \cite{mar01b}.}
\begin{minipage}{0.5\textwidth} 
\renewcommand{\footnoterule}{\kern -3pt} 
\begin{tabular}{llll}
\hline
 $\beta_{2}$ & $\beta_{3}$ & $\beta_{4}$ &
t$_{1/2}$ (y)\\ \hline
\\[-10pt]
1.2\footnote{$\epsilon_{2}$ and $\epsilon_{4}$ values for
$^{248}$Fm were taken from Ref. \cite{how80} and converted to
$\beta_{2}$ and $\beta_{4}$ values by extrapolation from the
curves given in Fig. 2 of Ref. \cite{naz96}.} &
0.0\footnote{Assuming $\beta_{3}$ = 0.}
 & 0.0$^{a}$ & 1.8$\times$10$^{11}$\\
1.2$^{a}$ & 0.19\footnote{Assuming $\beta_{3}$ = $\epsilon_{3}$ of
Ref. \cite{how80}.}
  & 0.0 & 4.6$\times$10$^{9}$\\
0.85\footnote{Parameters given in Ref. \cite{cwi94} for
$^{232}$Th.}
 & 0.35$^{d}$  & 0.18$^{d}$ & 1.3$\times$10$^{8}$\\
\hline
\end{tabular}
\end{minipage}
\renewcommand{\footnoterule}{\kern-3pt \hrule width .4\columnwidth
\kern 2.6pt}            
\vspace{-0.3cm}
\end{table}

\begin{figure}[h]
\includegraphics[width=0.48\textwidth]{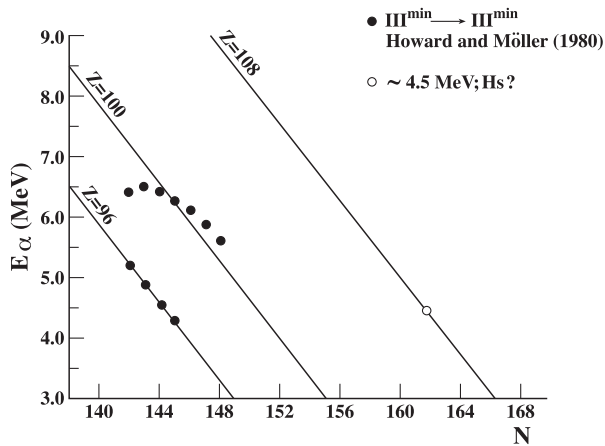}
\caption{Predictions \cite{how80}, and extrapolations for
III$^{min}$ $\rightarrow$ III$^{min}$ $\alpha$-particle energies.
The black dots are the predictions for various isotopes of Z=96
and Z=100. The straight lines are extrapolations from these
predictions. The open circle shows the position of 4.5 MeV
$\alpha$-particles in Z=108.}
\end{figure}

     Furthermore, in Fig. 2 deduced $\alpha$-particle energies for
     III$^{min}$ $\rightarrow$ III$^{min}$ transitions for some
     isotopes of Cm and Fm are presented. These energies were
     obtained from the predicted excitation energies \cite{how80}
     of the third minima in the relevant parent and daughter
     nuclei and from the known \cite{aud93} or predicted \cite{lir00} g.s. to
     g.s. Q$_\alpha$ values. The Q$_\alpha$ values
     for III$^{min}$ $\rightarrow$ III$^{min}$ transitions
       are much lower
     than the corresponding g.s. to g.s. values
     (see also Table 1 of Ref. \cite{mar01b}), since the excitation
      energy of the  III$^{min}$ in the parent nucleus is predicted
      to be at a lower value than in the daughter.
      An extrapolation of the deduced
      III$^{min}$ $\rightarrow$ III$^{min}$ transition energies to Z =
108 gives an E$_\alpha$ value of about 4.5 MeV  for N $\approx$
162. This is consistent with the suggestion \cite{mei70} that
$^{247}$Cm may be a descendent of an element with Z = 108 which
decays by the 4.5 MeV $\alpha$-particles, since $^{247}$Cm can be
obtained from $^{271}$Hs (N = 163) by six successive
$\alpha$-decays. Another possibility is that the long-lived
isotope is $^{267}$Hs which decays first by 4.5 MeV $\alpha$'s,
and then by two electron capture (EC) or $\beta^+$ decays and four
$\alpha$-decays to $^{247}$Cm. This scenario allows for a
transition from the third minimum to the first minimum because of
the total positive Q-value of the $\beta$(EC) decays of about 5
MeV \cite{mol97,kou00,lir00}. (Transitions from isomeric states to
normally deformed states by $\beta$(EC) decays have been seen
before \cite{mar87}.)\footnote{It should however be mentioned that
in principle the above 4.5 MeV $\alpha$-particles may  also be due
to a strongly retarded II$^{min}$ $\rightarrow$ I$^{min}$ or
III$^{min}$ $\rightarrow$ II$^{min}$ transition in the region of
Os itself. Such transitions have been seen in the Os - Ir - Hg
region \cite{mar01a} though with half-lives of several months.
(For normal 4.5 MeV $\alpha$-particles in Os the expected
\cite{vio66,roy00} half-life is about 1 y. Such short-lived
nuclide can not exist in nature).}

     Notwithstanding the uncertainty mentioned above about the origin of the giant
halos, let us now suggest a possible interpretation for these halos assuming that they are
due to $\alpha$-particles of around 10 and 13 MeV.
It has already been pointed out \cite{bra78,mol80}  that the
giant halos are associated with smaller halos. For instance, in the giant halo seen in
Fig.1 of Ref. \cite{gen70}, where the outer ring corresponds to $\alpha$-particles
of about 10 MeV, one
sees also a central black zone which could be due to low-energy $\alpha$-particles of 4.8 MeV.
A good candidate for the sequence of events producing this halo is a long-lived HD
isomeric state decaying by
a 4.8 MeV III$^{min}$ $\rightarrow$ III$^{min}$ $\alpha$-transition, followed by $\beta^+$(EC)
transitions to a normal state which decays by 10 MeV $\alpha$-particles. As a specific
example, one may consider the following scenario where a HD isomeric state in $^{282}$114
decays by 4.8 MeV $\alpha$'s to a HD isomeric state in $^{278}$112,
followed by two $\beta^+$(EC)
decays to a normal deformed state or to the g.s. of $^{278}$110. This latter nucleus is
predicted \cite{mol97,kou00,lir00} to decay by 10 MeV $\alpha$-particles.
For the deformation parameters given in
Table 1, the predicted T$_{1/2}$ value for a
4.8 MeV III$^{min}$ $\rightarrow$ III$^{min}$ $\alpha$-transition from $^{282}$114 is
10$^{8}$ - 10$^{11}$ y,
and the sum of the two Q$_{\beta}$ of
above 6 MeV \cite{mol97,kou00,lir00} makes the transition from
the isomeric state in the third minimum to a normal state in one of the daughter nuclei
possible.

     Finally, let us consider the giant halo presented in Fig. 3 of Ref. \cite{gen70}
      which would
correspond to 13.1 MeV $\alpha$-particles. Here too, in addition
to the large halo, one sees a small dark ring with an inner radius
which, if caused by $\alpha$-particles, correspond to a low energy
of about 5.1 MeV.\footnote{The widening of this ring is probably
due to an overexposure effect. Compare for instance Figs. 1l and
1i of Ref. \cite{gen74} where the halo from the 6.0-MeV
$\alpha$-group in Fig. 1l is quite wide.}
Similarly  to the situation above, one can propose as a scenario
for this halo a HD isomeric state in $^{316}$126 which decays by a
III$^{min}$ $\rightarrow$ III$^{min}$ low-energy
$\alpha$-transition to $^{312}$124, followed by two $\beta^+$(EC)
transitions, leading to the g.s. of $^{312}$122. The $^{312}$122
nucleus is predicted \cite{mol97,kou00} to decay by
$\alpha$-particles of around 13 MeV. For a 5.1 MeV HD to HD
$\alpha$-transition from $^{316}$126 the predicted \cite{mar01b}
half-life, using the parameters of Ref. \cite{cwi94} (Table 1,
last line) is 3x10$^{11}$ y. (Longer lifetimes are obtained for
the other sets of parameters in Table 1, whereas larger
deformation parameters, and consequently shorter lifetimes, may
exist in the nucleus $^{316}$126 compared to $^{232}$Th for which
the parameters in Ref. \cite{cwi94} were predicted.) Thus, a
consistent picture for the giant halo is possible if 5.1-MeV
$\alpha$-particles are the origin of the small halo within the
larger one.\footnote{Overexposure and blurring effects might be
the reasons why inner ring structure due to successive
disintegrations after the high energy $\alpha$-decays are not seen
in Figs. 1 and 3 of Ref. \cite{gen70}, since: ``Also in the Th-U
halos, where about 10$^{5}$ of them have been studied
\cite{gen74}, most have centers $>$ 20 microns diameter, which
generally produces halos without distinct inner ring structure,
whereas only relatively few possess the tiny, micron-size centers
necessary to produce halos with distinct rings." (Private
communication from R. V. Gentry).}

It should however be mentioned that the existence of low energy
$\alpha$-particle groups is not a necessary initial step for
observed giant halos. HD isomeric states which decay by
EC($\beta$) finally to the g.s. of superheavy nuclei, can explain
such giant halos as well.

Let us finally touch on the problem why these SD and HD isomeric
states which were used by us to explain the above phenomena do not
decay by fission.
 Their long lifetimes in the region
of 10$^{9}$ y, also against fission, are probably due to the
combined effect of the potential barrier and the high spin of the
state. As a matter of fact, back in 1969 \cite{nil69} a new type
of fission isomeric state has been predicted for nuclei with N
$\approx$ 144-150. A specialization energy \cite{whe56} in excess
of 4 MeV for the second barrier was predicted for a
[505]$\frac{11}{2}^{-}$ state, which is associated with a factor
of about 10$^{15}$ increase in the half-life of a normal fission
shape isomer.
\section{Possible Scenarios for Production of Superheavy Elements in Nature}
It is worthwhile mentioning that a natural way to produce HD high
spin isomeric states is by heavy ion reactions. First, high spin
states are produced by these reactions, and secondly, as seen in
Fig. 8 of Ref. \cite{mar01b}, for suitable projectile-target
combinations, the shape of the compound nucleus in the third
minimum fits with the shape of the projectile and target nuclei at
the touching point, and very little penetration and dissipation
energies  are involved in the reaction. The above isotopes
$^{267}$108, $^{282}$114 and $^{316}$126 could presumably be
produced by the following cold fusion reactions:

\begin{widetext}
a) $^{208}$Pb + $^{60}$Fe$ \rightarrow$ $^{267}$Hs + n
(Q$_{value}$ = -214 MeV; C.B. = 223 MeV)

b)  $^{208}$Pb + $^{74}$Ge $\rightarrow$ $^{282}$114   (Q$_{value}$ = -262 MeV;
C.B. = 267 MeV)

c)  $^{232}$Th + $^{84}$Kr $\rightarrow$ $^{316}$126  (Q$_{value}$ = -319 MeV;
C.B. = 317 MeV)

d)  $^{238}$U + $^{78}$Se $\rightarrow$  $^{316}$126 (Q$_{value}$
= -302 MeV; C.B. = 307 MeV),\\
\end{widetext}

where the projectile and target nuclei are stable or quasi-stable
isotopes. (C.B. is the Coulomb barrier between the projectile and
the target nucleus for a radius parameter R$_{0}$ = 1.4 fm).
\section{Summary}
In summary, it has been shown that the newly discovered long-lived
super- and hyperdeformed isomeric states and their unusual
radioactive decay properties enable one to understand certain
previously puzzling phenomena seen in nature. Thus, the Po halos
can be understood as being due to the existence of such isomeric
states in nuclei around $^{210}$Po, $^{214}$Po and $^{218}$Po
which undergo $\beta$- and $\gamma$-decays to the ground states of
these isotopes. Likewise, the low-energy enhanced 4.5 MeV
$\alpha$-particle group \cite{cher63,chery64,cher68,mei70} can be
quantitatively understood as a hyperdeformed to hyperdeformed
transition from an isotope with Z = 108, A $\approx$ 270. Finally,
it was shown that the giant halos can be consistently interpreted
as being due to 10 and 13 MeV $\alpha$-particles following low
energy III$^{min}$ $\rightarrow$ III$^{min}$ transitions in
superheavy nuclei around Z = 114 and Z = 126, respectively, which
eventually decay to normal states emitting such high-energy
$\alpha$-particles. The existence in both cases of halos with
small radii, which might be related to hyperdeformed to
hyperdeformed transitions, along with the large halos, lends
support to this scenario. It has however been pointed out that the
existence of low energy $\alpha$-particle groups as an initial
step is not a necessary condition for the interpretation of the
giant halos. III$^{min}$ isomeric states which decay by
EC($\beta$) and $\gamma$-rays finally to the ground states of
superheavy nuclei, would lead to giant halos as well.

Based on the above it seems to us that the search for superheavy
elements in nature should be pursued.
\section{Acknowledgements}
We appreciate valuable discussions with R. V. Gentry, J. L. Weil
and N. Zeldes. D. K. acknowledges financial support from the
DFG.\\

\end{document}